\documentclass[12pt]{article}
\usepackage{latexsym}   
\usepackage{amsmath,  amssymb,  amsthm,  latexsym}

\newcommand{\rw}{\rightarrow}

\newcommand{\ds}{\delta^{\ast}}
\newcommand{\sa}{\Sigma^{\ast}}
\newcommand{\es}{\eta^{\ast}}

\begin{document}
\newtheorem{lemma}{Lemma}[section]
\newtheorem{proposition}{Proposition}[section]
\newtheorem{theorem}{Theorem}[section]
\newtheorem{corollary}{Corollary}[section]
\newtheorem{example}{Example}[section]
\newtheorem{definition}{Definition}[section]
\newtheorem{remark}{Remark}[section]
\baselineskip 28pt

\title{\Large\bf Finite Automata Based on Quantum Logic and Their Determinization
\thanks{This work is supported by National Science Foundation of China
(Grant No.10571112),  National 973 Foundation Research Program
(Grant No.2002CB312200) and Key Research Project of Ministry of
Education of China (No.107106).}}
\author{{Yongming Li}\\
  {\small College of Computer Science,}
  {\small Shaanxi Normal University, Xi'an, 710062, China}\\
  {\small Email:liyongm@snnu.edu.cn, Tel.: +862985310166, Fax:
  +862985310161}}

\date{}
\maketitle
\begin{center}
\begin{minipage}{140mm}
\centerline{\bf Abstract} \vskip 3mm {We give the quantum subset
construction of orthomodular lattice-valued finite automata, then we
show the equivalence between orthomodular lattice-valued finite
automata, orthomodular lattice-valued deterministic finite automata
and orthomodular lattice-valued finite automata with empty
string-moves. Based on these equivalences, we study the algebraic
operations on orthomodular lattice-valued regular languages, then we
establish Kleene theorem in the frame of quantum logic.}


\end{minipage}
\end{center}

\vskip 3mm

\baselineskip 20pt

\section{Introduction}

In classical computation theory, characterizing all formal
languages, or even better sorting them in some hierarchy, was an
important issue. For example, the most restricted class-the regular
languages-can be characterized by finite automata and by regular
expressions. It is well-known that regular languages can be
recognized by deterministic finite automata, nondeterministic finite
automata (with or without empty string-moves), the technique to
prove the equivalence between nondeterministic and deterministic
automata is the subset construction
\cite{rabin59,hopcroft79,khoussainov01}. Another important result in
classical automata theory is the Kleene theorem which shows the
equivalence between finite automata and regular expressions. All
these results have been extended to fuzzy finite automata as the
fuzzy computing models by introducing fuzzy subset construction and
fuzzy regular expressions, see \cite{li05} for the detail. In the
frame of weighted automata theory, the subset construction and
Kleene theorem have been also discussed, see \cite{esik07} for the
detail. The subset construction and Kleene theorem form the basic
results for the computational models with different purposes. From
this core the theory has developed into numerous directions.
Computing models of quantum computation is a new research along this
direction.

The ideas of quantum computing came from the connections between
physics and computation \cite{nc00}. In particular, in 1994 Shor
discovered a polynomial-time algorithm for factoring prime on
quantum computers, and Grover then found an algorithm for searching
through a database in square root time. Since then, quantum
computing has attracted more and more attention in the research
community. In this field, the computing models of quantum
computation is still one of the most important topic to study.
Quantum finite automata can be viewed as a kind of quantum computer
model with finite memory, for which we may refer to ref.
\cite{gruska99,moore00,ambainis02}. A more fundamental issue
regarding quantum computing models may be automata theory based on
quantum logic \cite{ying001,ying002,ying05,qiu04,qiu071,qiu072}
 (called orthomodular lattice-valued automata).
Quantum logic was suggested by Birkhoff and Neumann in 1936 for
studying the logical basic of quantum mechanics, and it originated
from the Hilbert space's formalization of quantum mechanics. Since a
state of a quantum system can be described by a closed subspace of a
Hilbert space, while all closed subspaces of a Hilbert space are
endowed with the algebraic structure of orthomodular lattices, it
was proposed that orthomodular lattices were thought of as the
algebraic version of quantum logic. Actually, orthomodular lattices
sometimes are defined directly as quantum logic. Thus, investigating
orthomodular lattice-valued automata may be considered to be an
important aspect of the logical basic of quantum computing.
Recently, the author \cite{ying001,ying002,ying05} primarily and
very significantly considered automata theory based on quantum logic
($l$-valued automata), in which quantum logic is understood as a
logic whose truth-value set is an orthomodular lattice, and an
element of an orthomodular lattice is assigned to each transition of
an automaton and it is considered to be the truth value of the
proposition describing the transition. This is a logical approach to
quantum computation in which the ultimately objective is to manage
to set up the logic platform for the quantum computation, and it
should be treated as a further abstraction of mathematical models of
quantum computation. With this approach, the author dealt with some
operations on $l$-valued automata, and interestingly established
corresponding pumping lemma, showed the equivalence between the
distributivity of truth-value lattices and the product operation of
orthomodular lattice-valued automata, etc., showed an essential
difference exists between the classical theory of computation and
the computation theory based on quantum logic.

The concept of an orthomodular lattice-valued finite automaton is a
natural generalization of the concept of a nondeterministic
automaton, as the concepts of an orthomodular lattice-valued set and
an orthomodular lattice-valued relation are generalizations of the
classical concepts of a set and a relation. Relationships between
orthomodular lattice-valued nondeterministic and deterministic
automata have been studied by Ying \cite{ying05}. The method for
determinization of  orthomodular lattice-valued automata used by
Ying in \cite{ying05} is analogous to the well-known subset
construction, and it is called here the extended subset
construction. Unfortunately, extended subset construction does not
work well for orthomodular lattice-valued finite automaton. As shown
by Ying in \cite{ying05}, under extended subset construction, one
can not prove the equivalence between orthomodular lattice-valued
nondeterministic and deterministic automata. In fact, Ying proved
that the equivalence between orthomodular lattice-valued
nondeterministic and deterministic automata under extended subset
construction is equivalent to the underlying logic being classical
logic (i.e., the used truth structure as an orthomodular lattice
must be a Boolean algebra). It is left open as a problem whether
orthomodular lattice-valued nondeterministic and deterministic
automata are equivalent. We shall introduce quantum subset
construction in this paper to study this problem. Indeed, using the
quantum subset construction introduced in this paper, we show the
equivalence between orthomodular lattice-valued nondeterministic
(with or without empty string-moves) and deterministic automata
. Furthermore, we characterize the quantum languages recognized by
orthomodular lattice-valued automata by the orthomodular
lattice-valued recognizable step languages in Theorem
\ref{thm:regular}, which have very simple construction. Using this
characterization of recognizable quantum languages, we further show
the Kleene theorem holds in the frame of quantum logic. Many results
in \cite{ying05} can be strengthen in this manner.

The content of this paper is arranged as follows. In Section 2, we
first recall the definition of orthomodular lattice-valued automata,
then we introduce the notion of orthomodular lattice-valued
deterministic automata. By introducing the quantum subset
construction, we prove the equivalence between orthomodular
lattice-valued nondeterministic (with or without empty string-moves)
automata and deterministic automata. In Section 3, we first give a
simple characterization of quantum regular languages, the operations
property of quantum regular languages is discussed. Then the Kleene
theorem in quantum logic is presented. Some conclusion is presented
finally.

\section  {Determinization of $l$-valued finite automata and quantum subset constructions}

Quantum logic is understood as a (complete) orthomodular
lattice-valued logic, for the detail, we refer to
\cite{kalmbach83,qiu072,ying05}. We briefly recall some notions and
notations of quantum logic. An ortholattice is a 7-tuple
$l=(L,\leq,\wedge,\vee,\bot,0,1)$, where
$l=(L,\leq,\wedge,\vee,\bot,0,1)$ is a bounded lattice,  0 and 1 are
the least and largest elements of $L$, respectively, $\leq$ is the
partial ordering in $L$; and for any $a,b\in L$, $a\wedge b$ and
$a\vee b$ stand for the greatest lower bound (or meet) and the least
upper bound (or join) of $a$ and $b$, respectively. $\bot$ is a
unary operation on $L$, called orthocomplement, and required to
satisfy the following conditions: for any $a,b\in L$, $a\wedge
a^{\bot}=0$, $a\vee a^{\bot}=1$; $a^{\bot\bot}=a$; $a\leq b$ implies
$b^{\bot}\leq a^{\bot}$. An orthomodular lattice is an ortholattice
$l=(L,\leq,\wedge,\vee,\bot,0,1)$ satisfying the orthomodular law:
for all $a,b\in L$, $a\leq b$ implies  $a\wedge (a^{\bot}\vee b)=b$.
A quantum logic is a (complete) orthomodular lattice-valued logic
(called $l$-valued logic). Defined an implication operator $\rw$ on
$l$ satisfying: for all $a,b\in L$, $a\leq b$ if and only if (iff)
$a\rw b=1$. In this paper, we use Sasaki arrow as the implication
operator. Sasaki arrow is defined as follows: for all $a,b\in L$,
$a\rw b=a^{\bot}\vee(a\wedge b)$. The bi-implication operator
$\leftrightarrow$ is defined as follows: for all $a,b\in L$,
$a\leftrightarrow b=(a\rw b)\wedge (b\rw a)$. The syntax of
$l$-valued logic is similar to that of classical first-order logic.
We have three primitive connectives $\neg$ (negation), $\vee$
(conjunction) and $\rw$ (implication), and a primitive quantifier
$\exists$ (existential quantifier). The connectives $\wedge$
(conjunction) and $\leftrightarrow$ (bi-implication) and the
universal quantifier $\forall$ are defined in terms of $\neg$,
$\vee$, $\rw$ and $\exists$ in the usual way. In addition, we need
to use some set-theoretical formulas. Let $\in$ (membership) be a
binary (primitive) predicate symbol. Then $\subseteq$ and $\equiv$
(equality) can be defined with $\in$ as usual. The semantics of
$l$-valued logic is given by interpreting the connectives $\neg$,
$\vee$ and $\rw$ as the operations $\bot$, $\vee$ and $\rw$ on $L$,
respectively, and interpreting the quantifier $\exists$ as the least
upper bound in $l$. Moreover, the truth value of set-theoretical
formula $x\in A$ is $[x\in A]=A(x)$. In the $l$-valued logic, 1 is
the unique designated truth value; a formula $\varphi$ is valid iff
$[\varphi]=1$, and denoted by $\models_l\varphi$. For a finite
subset $X$ of $l$, the (commutator) $\gamma(X)$ generated by $X$ is
defined as follows:

$\gamma(X)=\vee\{\wedge_{a\in X}a^{f(a)}: f: X\rw \{1,-1\}$ is a
mapping$\}$,

\noindent where, $x^1=x$, $x^{-1}=x^{\bot}$.

In order to distinguish the symbols representing languages and the
symbols representing lattices, we use symbol $l$ to represent
orthomodular lattice, and use $L$ to represent language. We use the
symbols $a,b,c,d,k$ to represent the elements of $l$.

\begin{definition}\label{def:l-vfa}{\rm \cite{ying05}}
{\rm An {\sl $l$-valued finite automaton} ($l$-VFA for short) is a
5-tuple ${\cal A}=(Q,\Sigma,\delta,I,F)$, where $Q$ denotes a finite
set of states, $\Sigma$ a finite input alphabet, and $\delta$ is an
$l$-valued subset of $Q\times \Sigma\times Q$; that is, a mapping
from $Q\times \Sigma\times Q$ into $l$, and it is called the
$l$-valued (quantum) transition relation. Intuitively, $\delta$ is
an $l$-valued (ternary) predicate over $Q$, $\Sigma$ and $Q$, and
for any $p,q\in Q$ and $\sigma\in \Sigma$, $\delta(p,\sigma,q)$
stands for the the truth value (in quantum logic) of the proposition
that input $\sigma$ causes state $p$ to become $q$. $I$ and $F$ are
$l$-valued subset of $Q$; that is, a mapping form $Q$ into $l$,
which represent the initial state and final states, respectively.
For each $q\in Q$, $I(q)$ indicates the truth value (in the
underlying quantum logic) of the proposition that $q$ is an initial
state, $F(q)$ expresses the truth value (in our quantum logic) of
the proposition
 that $q$ is a finial state.}

\end{definition}

The propositions of the form

``$q$ is an initial state", written ``$q\in I$''.

``$q$ is a final state", written ``$q\in F$".

``input $\sigma$ causes state $q$ to become $p$, according to the specification given by $\delta$"
, written ``$(q,\sigma,p)\in \delta$".

\noindent denote the atomic propositions in our logical languages
designated for describing $l$-valued automaton ${\cal A}$. The truth
values of the above three propositions are respectively $I(q)$,
$F(q)$ and $\delta(q,\sigma,p)$. We use the symbols $\sigma,\tau$ to
represent the elements in $\Sigma$, use the symbols $\omega,\theta$
to denote the strings over $\Sigma$,  and use $\varepsilon$ to
represent the empty string over $\Sigma$. We use the symbols ${\cal
A}, {\cal B}$ to denote the $l$-valued finite automata.

For an $l$-VFA ${\cal A}$, the {\sl $l$-valued unary recognizability
predicate} $rec_{\cal A}$ over $\sa$ is defined as a mapping from
$\sa$ into $l$: for each $\omega\in \sa$, let $\omega
=\sigma_1\cdots \sigma_n$ for some $n\geq 0$,

$rec_{\cal A}(\omega)=(\exists q_0\in Q)\cdots(\exists q_n\in Q$).$(q_0\in I\wedge q_n\in F\wedge
(q_0,\sigma_1, q_1)\in \delta\wedge\cdots \wedge(q_{n-1},\sigma_n, q_n)\in \delta)$.

In other words, the truth value of the proposition that $\omega$ is
recognizable by ${\cal A}$ is given by

$[rec_{\cal A}(\omega)]=\bigvee\{I(q_0)\wedge\delta(q_0,
\sigma_1,q_1)\wedge\cdots\wedge \delta(q_{n-1},\sigma_n,q_n)\wedge
F(q_n): q_0, \cdots, q_n\in Q\}$.

We call $rec_{\cal A}$ the $l$-valued language recognized or
accepted by $l$-VFA ${\cal A}$. We use $l(\sa)$ to denote the set of
all $l$-valued language over $\sa$, which is an $l$-valued subset of
$\sa$; that is, a mapping from $\sa$ to $l$. We also call $l$-valued
languages by quantum languages. For an $A\in l(\sa)$, if there is an
$l$-VFA ${\cal A}$ such that $A=rec_{\cal A}$, then we call $A$ an
{\sl $l$-valued regular language} or {\sl $l$-regular language} on
$\Sigma$, which is also called {\sl quantum regular language}
without mentioned the truth-valued lattice.

Furthermore, we can define unary predicate $Rec_{\Sigma}$ on
$l(\sa)$ as follows: for all $B\in l(\sa)$,

$Rec_{\Sigma}(B)=(\exists {\cal A}\in \mathbf{A}$ $(\Sigma)).
(B\equiv rec_{\cal A})$.

\noindent where $\mathbf{A}$ $(\Sigma)$ writes for the class of all
$l$-valued automata over $\Sigma$, we refer to \cite{ying05} for the
detail.



First, we show that the image set of each quantum regular langauge is always a finite set of $l$.

\begin{lemma}\label{le:image}{\rm \cite{li93}}
Let $l$ be a lattice, and $X$ a finite subset of $l$. Then the
$\wedge$-semilattice of \ $l$ generated by $X$, written as
$X_{\wedge}$, is finite, the $\vee$-semilattice of \ $l$ generated
by $X$, denoted $X_{\vee}$, is also finite, where
$X_{\wedge}=\{x_1\wedge \cdots\wedge x_k: k\geq 1, x_1,\cdots,
x_k\in X\}\cup \{1\}$, and $X_{\vee}=\{x_1\vee \cdots\vee x_k: k\geq
1, x_1,\cdots, x_k\in X\}\cup \{0\}$.

\end{lemma}

\begin{proposition}\label{pro:image}

Let ${\cal A}=(Q, \Sigma, \delta, I, F)$ be an $l$-VFA. Then the
image set of the quantum language $rec_{{\cal A}}$, as a mapping
from $\sa$ to \ $l$, is finite; that is, the subset $Im(rec_{{\cal
A}})= \{ r\in l: \exists \omega\in \sa, [rec_{{\cal
A}}(\omega)]=r\}$ of \ $l$ is finite.

\end{proposition}

\noindent {\bf Proof}\ \ For any $\omega=\sigma_1\cdots \sigma_k\in
\sa$, observing that $[rec_{{\cal A}}(\omega)]=\bigvee\{I(q_0)\wedge
\delta(q_0,\sigma_1,q_1)\wedge\cdots\wedge
\delta(q_{k-1},\sigma_k,q_k)\wedge F(q_k): q_0,\cdots, q_k\in Q\}$.
On input $\omega=\sigma_1\cdots \sigma_k\in \sa$, there are only
finite accepting paths, assumed as $m$, causing an initial state
$q_0\in I$ to become a final state $q_k\in F$. For the $i$-th
accepting path, we let $a_{i0}=I(q_0)$,
$a_{i1}=\delta(q_0,\sigma_1,q_1)$, $\cdots,
a_{ik}=\delta(q_{k-1},\sigma_k,q_k)$ and $a_{i,k+1}=F(q_k)$. Then
the truth value of $rec_{\cal A}(\omega)$ can be calculated as,
$[rec_{\cal A}(\omega)]=(a_{10}\wedge\cdots\wedge a_{1k}\wedge
a_{1,k+1})\vee\cdots\vee (a_{m0}\wedge\cdots\wedge a_{mk}\wedge
a_{m,k+1})$. Let $X=Im(\delta)\cup Im(I)\cup Im(F)$, then $X$ is
obvious a finite subset of $l$ and $a_{ij}\in X$ for any $1\leq
i\leq m$ and $0\leq j\leq k+1$. For any $\omega\in \sa$, by the
above observation, we know that $[rec_{{\cal A}}(\omega)]\in
(X_{\wedge})_{\vee}$, so $Im(rec_{{\cal A}})\subseteq
(X_{\wedge})_{\vee}$. By Lemma \ref{le:image}, $(X_{\wedge})_{\vee}$
is a finite subset of $l$, and thus $Im(rec_{{\cal A}})$, as a
subset of $(X_{\wedge})_{\vee}$, is also a finite subset of $l$.
\hfill$\Box$

Due to Proposition \ref{pro:image}, for any $l$-VFA, the image set
of its recognizable quantum language is always finite. Then we have
the following observation: the orthomodular lattice $l$ may be
infinite as a set, but for a given $l$-VFA ${\cal A}$, only a finite
subset of $l$ is employed in the operating of ${\cal A}$. This
observation is the core in the introducing of quantum subset
construction in this section.

The notion of nondeterminism plays a central role in the theory of
computation. The nondeterministic mechanism enables a device to
change its states in a way that is only partially determined by the
current state and the input symbol. The concept of $l$-VFA is
obviously a generalization of nondeterministic finite automaton (NFA
for short). In classical theory of automata, each nondeterministic
finite automaton is equivalent to a deterministic one; more
precisely, there exists a deterministic finite automaton (DFA for
short) which accepts the same language as the originally given
nondeterministic one does. The construction of DFA from an NFA is
the well-know subset construction introduced by Rabin and Scott
\cite{rabin59}. With respect to the case of $l$-VFA, the situation
is more complex. In fact, as shown in \cite{ying05}, the subset
construction does not work well for $l$-VFA. That is, for an $l$-VFA
${\cal A}$, one can construct an $l$-valued deterministic finite
automaton ${\cal B}$, as defined in \cite{ying05} using the subset
construction. However, ${\cal B}$ is not necessarily equivalent to
${\cal A}$, i.e., the equality $rec_{\cal A}=rec_{\cal B}$ does not
hold in general. Some conditions that guarantee the equivalence
between ${\cal A}$ and ${\cal B}$ are given in \cite{ying05}.
Therefore, it is an open problem whether an $l$-VFA can always be
determinizable. We shall show that the answer is affirmative. We
shall introduce subset construction in the frame of quantum logic
which we call it the quantum subset construction. First, we define a
new kind of deterministic $l$-VFA, which is stronger than that given
in \cite{ying05} using the same name. We require some stronger
condition for the quantum transition.

\begin{definition}\label{def:l-dfa}

{\rm An {\sl $l$-valued deterministic finite automaton} ($l$-VDFA
for short) is a 5-tuple ${\cal A}=(Q,\Sigma,\delta,q_0,F)$, where
$Q$, $\Sigma$ and $F$ are the same as in an $l$-valued automaton,
$q_0\in Q$ is the initial state, and the quantum transition relation
$\delta$ is crisp and deterministic; that is, $\delta$ is a mapping
from $Q\times \Sigma$ into $Q$.}

\end{definition}

Note that our definition differs from the usual definition of a
deterministic automaton only in that the final states form an
$l$-valued subset of $Q$. This, however, makes it possible to accept
words to certain truth degrees (in the underlying quantum logic),
and thus to recognize quantum languages.

For an $l$-VDFA, ${\cal A}=(Q,\Sigma,\delta,q_0,F)$, its
corresponding $l$- valued recognizability predicate $rec_{\cal A}\in
l(\sa)$ is defined as: for all $\omega=\sigma_1\cdots\sigma_n\in
\sa$,

$rec_{\cal A}(\omega)=(\exists q_1\in Q)\cdots(\exists q_n\in Q$).$(q_n\in F\wedge
\delta(q_0,\sigma_1)=q_1\wedge\cdots \wedge \delta(q_{n-1},\sigma_n)=q_n)$.

Write $\ds$ the extension of transition relation $\delta$ by putting $\ds(q,\varepsilon)=q$
and $\ds(q,\omega\sigma)=\delta(\ds(q,\omega),\sigma)$ for any $q\in Q$ and
$\omega\in\sa$ and $\sigma\in \Sigma$,
then the truth value of the proposition $rec_{\cal A}(\omega)$ is given by,

$[rec_{\cal A}(\omega)]=F(\ds(q_0, \omega))$.

Obviously, the notion of $l$-VDFA is a special case of $l$-valued deterministic
automata defined in \cite{ying05}, but the converse inclusion does not hold in general.

For any $l$-VFA, ${\cal A}=(Q,\Sigma,\delta,q_0,F)$, we now
introduce the {\sl quantum subset construction} to construct an
equivalent $l$-VDFA ${\cal A}^d=(Q^d, \Sigma, \eta, S, E)$ from
${\cal A}$.

Let $X=Im(\delta)\cup Im(I)\cup Im(F)$, then $X$ is obvious a finite
subset of $l$. Let $l_1=X_{\wedge}$. By Lemma \ref{le:image}, $l_1$
is a $\wedge$-semilattice of $l$ generated by $X$ and is also finite
subset of $l$. Choose

$Q^d=2^{Q\times (l_1-\{0\})}$,

\noindent where $2^{Q\times (l_1-\{0\})}$ denotes the set of all
subsets of $Q\times (l_1-\{0\})$. Then $Q^d$ is obvious a finite
set. Take

$S=\{(q,I(q)): q\in Q$ and $I(q)\not=0\}$,

\noindent then $S\in Q^d$. The state transition relation $\eta:
Q^d\times \Sigma\rw Q^d$ is defined as, for any $(q,r)\in Q\times
(l_1-\{0\})$ and $\sigma\in \Sigma$,

$\eta(\{(q,r)\},\sigma)=\{(p,\delta(q,\sigma,p)\wedge r): p\in Q$
and $\delta(q,\sigma, p)\wedge r\not=0\}$,

\noindent and for $Z\in Q^d$,

$\eta(Z,\sigma)=\bigcup\{\eta(\{(q,r)\},\sigma): (q,r)\in Z\}$.

\noindent By the definition of $l_1$,  $l_1$ is closed under finite
meet operation, i.e., for any $a,b\in l_1$, $a\wedge b\in l_1$, it
follows that, for any $r\in l_1$ and for any $(p,\sigma,q)\in
Q\times \Sigma\times Q$, $r\wedge \delta(p,\sigma,q)\in l_1$, and
thus $\eta(\{(q,r)\},\sigma)\in Q^d$ for any $(q,r)\in Q\times
(l_1-\{0\})$. Then the mapping $\eta$ is well defined. The
$l$-valued final state $E: Q^d\rw l$ is defined by, for any $Z\in
Q^d$,

$E(Z)= \bigvee\{r\wedge F(q): (q,r)\in Z\}$.

\noindent Then ${\cal A}^d$ is an $l$-VDFA.

\begin{theorem}\label{th:determinization}

For any $l$-VFA, ${\cal A}=(Q,\Sigma,\delta,q_0,F)$, the $l$-VDFA
${\cal A}^d=(Q^d, \Sigma, \eta, S, E)$ constructed above is
equivalent to ${\cal A}$, i.e., $rec_{\cal A}=rec_{{\cal A}^d}$. In
the language of quantum logic, it means that, for any $\omega\in
\sa$, $$\models_l rec_{\cal A}(\omega)\leftrightarrow rec_{{\cal
A}^d}(\omega).$$

\end{theorem}

\noindent {\bf Proof}\ \ We wish to show by induction on the length
$|\omega|$ of input string $\omega$ that
$\eta^{\ast}(S,\omega)=\{(q_n,I(q_0)\wedge\delta(q_0,\sigma_1,q_1)
\wedge\cdots\wedge\delta(q_{n-1},\sigma_n,q_n)): q_0,\cdots,q_n\in
Q$ and $r_n=I(q_0)\wedge\delta(q_0,\sigma_1,q_1)
\wedge\cdots\wedge\delta(q_{n-1},\sigma_n,q_n)\not=0\}$, where
$\omega=\sigma_1\cdots\sigma_n$ for $n\geq 0$. The results is
trivial for $|\omega|=0$, since $\omega=\varepsilon$ and
$\eta^{\ast}(S,\omega)=\{(q_0,I(q_0)): q_0\in Q$ and
$I(q_0)\not=0\}$. Suppose that the hypothesis is true for inputs of
length $n$ or less. Let $\omega=\sigma_1\cdots\sigma_{n+1}$ be a
string of length $n+1$, write $x=\sigma_1\cdots\sigma_n$, then
$\omega=x\sigma_{n+1}$. Then

$\eta^{\ast}(S,x\sigma_{n+1})=\eta(\eta^{\ast}(S,x),\sigma_{n+1})$.

By the inductive hypothesis,

$\eta^{\ast}(S,x)=\{(q_n,I(q_0)\wedge\delta(q_0,\sigma_1,q_1)
\wedge\cdots\wedge\delta(q_{n-1},\sigma_n,q_n)): q_0,\cdots,q_n\in
Q$ and $r_n=I(q_0)\wedge\delta(q_0,\sigma_1,q_1)
\wedge\cdots\wedge\delta(q_{n-1},\sigma_n,q_n)\not=0\}$.

By the definition of $\eta$,

$\eta(\eta^{\ast}(S,x),\sigma_{n+1})$=$\bigcup_{(q_n,r_n)\in
\eta^{\ast}(S,x)}\eta(\{(q_n,r_n)\},\sigma_{n+1})$=$\bigcup_{(q_n,r_n)\in
\eta^{\ast}(S,x)}\{(q_{n+1},r_n\wedge
\delta(q_n,\sigma_{n+1},q_{n+1}))$: $q_{n+1}\in Q$ and $r_n\wedge
\delta(q_n,\sigma_{n+1},q_{n+1})\not=0\}=\{(q_{n+1},I(q_0)\wedge\delta(q_0,\sigma_1,q_1)
\wedge\cdots\wedge\delta(q_{n-1},\sigma_n,q_n)\wedge
\delta(q_n,\sigma_{n+1},q_{n+1})): q_0,\cdots,q_{n+1}\in Q$ and
$r_{n+1}=I(q_0)\wedge\delta(q_0,\sigma_1,q_1)
\wedge\cdots\wedge\delta(q_{n},\sigma_n,q_{n+1})\not=0\}$

\noindent which establishes the inductive hypothesis.

By the definition of $l$-valued final state $E$, for any input
$\omega=\sigma_1\cdots\sigma_n\in\sa(n\geq 0)$, we have

$[rec_{{\cal
A}^d}(\omega)]=E(\eta^{\ast}(S,\omega))=\bigvee\{r_n\wedge F(q_n):
(q_n,r_n)\in
\eta^{\ast}(S,\omega)\}=\bigvee\{I(q_0)\wedge\delta(q_0,\sigma_1,q_1)
\wedge\cdots\wedge\delta(q_{n-1},\sigma_n,q_n)\wedge
F(q_n):q_0,\cdots, q_n\in Q$ and
$I(q_0)\wedge\delta(q_0,\sigma_1,q_1)
\wedge\cdots\wedge\delta(q_{n-1},\sigma_n,q_n)\not=0\}
=\bigvee\{I(q_0)\wedge\delta(q_0,\sigma_1,q_1)
\wedge\cdots\wedge\delta(q_{n-1},\sigma_n,q_n)\wedge
F(q_n):q_0,\cdots, q_n\in Q\}=[rec_{\cal A}(\omega)]$. Thus
$rec_{{\cal A}^d}=rec_{\cal A}$, ${\cal A}$ and ${\cal A}^d$ are
equivalent. \hfill$\Box$

Theorem \ref{th:determinization} gives the subset construction of
finite automaton in the frame of quantum logic. In fact, in the case
of $l=\{0,1\}$, the underlying logic is the classical logic, the
quantum subset construction is just the ordinary subset
construction.

We give an example to illustrate the technique of the quantum subset
construction introduced above.

\begin{example}\label{ex:detreminization}

{\rm Let $\otimes^2\mathbb{C}^2$ be the 2-qubit state space, where
$\mathbb{C}$ denotes the set of complex numbers. All the closed
subspaces of Hilbert space $\otimes^2\mathbb{C}^2$, denoted by $l$,
forms a (complete) orthomodular lattice (\cite{kalmbach83}), $(l,
\leq, \wedge,\vee,\bot,0,1)$, with usual notations. As the standard
notation in quantum computation (\cite{gruska99,nc00}), $\mid
0\rangle \mid 0\rangle$, $\mid 0\rangle \mid 1\rangle$, $\mid
1\rangle \mid 0\rangle$, $\mid 1\rangle \mid 1\rangle$ are four
basis states in the 2-qubit state space. We use $a_{ij}=span(\mid
i\rangle \mid j\rangle)$ to denote the closed subspace spanned by
$\mid i\rangle \mid j\rangle$, $i,j=0,1$.

An $l$-VFA ${\cal A}=(Q,\Sigma,\delta,I,F)$ is defined as follows
(c.f., \cite{qiu071}), $Q=\{p,q\}$, $\Sigma=\{\sigma\}$, $I(p)=1$
and $I(q)=a_{10}$, $F(p)=a_{10}$ and $F(q)=1$, and
$\delta(p,\sigma,q)=a_{00}, \delta(p,\sigma,p)=a_{01},
\delta(q,\sigma,q)=a_{10}$ and $\delta(q,\sigma,p)=a_{11}$.

Using the quantum subset construction, the determinization of ${\cal
A}$ is induced as follows. In this example, $l$ is an infinite
orthomodular lattice, and $l_1=\{a_{00},a_{01},a_{10},a_{11},0,1\}$.
In the construction of ${\cal A}^d$, the state set $Q^d$ is
$2^{Q\times (l_1-\{0\})}$, ${\cal A}^d$ will have $2^{10}$ states.
To give a full construction of ${\cal A}^d$ is a tedious work.
However, it is sufficient to give those states which are useful in
generating the $l$-valued language recognized by ${\cal A}^d$ from
the initial state $S$.

The initial state is $S=\{(p,1),(q,a_{10})\}$. By the simple
calculation, we have
$\eta(S,\sigma)=\{(p,a_{01}),(q,a_{00}),(q,a_{10})\}$,
$\eta(\{(p,a_{01}),(q,a_{00}),(q,a_{10})\},\sigma)=\{(p,a_{01}),(q,a_{10})\}$,
and $\eta(\{(p,a_{01}),(q$,
$a_{10})\},\sigma)=\{(p,a_{01}),(q,a_{10})\}$. Therefore, the useful
states of ${\cal A}^d$ are $S, \{(p,a_{01}),(q,a_{00}),(q,a_{10})\}$
and $\{(p,a_{01}),(q,a_{10})\}$, which are denoted as $p_0,p_1$ and
$p_2$ respectively. Let $P=\{p_0,p_1,p_2\}$, then the state
transition function $\eta$ is defined as, $\eta(p_0,\sigma)=p_1,
\eta(p_1,\sigma)=p_2$ and $\eta(p_2,\sigma)=p_2$. The $l$-valued
final state $E$ is defined as, $E(p_0)=(1\wedge F(p))\vee
(a_{10}\wedge F(q))=a_{01}\vee a_{10}$, $E(p_1)=(a_{01}\wedge
F(p))\vee (a_{00}\wedge F(q))\vee (a_{10}\wedge F(q))=a_{01}\vee
a_{00}\vee a_{10}$ and $E(p_2)=(a_{01}\wedge F(p))\vee (a_{10}\wedge
F(q))=a_{01}\vee a_{10}$. This complete the construction of ${\cal
A}^d=(P,\Sigma,\eta,p_0,E)$. Then $rec_{\cal A}(=rec_{{\cal A}^d})$
can be simply calculated as follows,

\[[rec_{\cal A}(\omega)] =\left\{
\begin{array}{ll}
a_{01}\vee a_{00}\vee a_{10}, & $ if $ \omega=\sigma,\\
a_{01}\vee a_{10}, & $ otherwise$ .
\end{array}
\right.\] }

\end{example}

We continue to study the relationship between $l$-VFA and $l$-VFA
with $\varepsilon$-moves. Let us first recall the definition of
$l$-VFA with $\varepsilon$-moves.

\begin{definition}\label{def:e-moves}{\rm \cite{ying05}}
{\rm An {\sl $l$-valued automaton with $\varepsilon$-moves}
($l$-VFA$_{\varepsilon}$ for short) is a five-tuple ${\cal
A}=(Q,\Sigma,\delta,I,F)$ in which all components are the same as in
an $l$-valued automaton (without $\varepsilon$-moves), but the
domain of the quantum transition relation $\delta$ is changed to
$Q\times (\Sigma\cup\{\varepsilon\})\times Q$; that is, $\delta$ is
a mapping from $Q\times (\Sigma\cup\{\varepsilon\})\times Q$ into
$l$, where $\varepsilon$ stands for the empty string of input
symbols.}

\end{definition}

Now let ${\cal A}=(Q,\Sigma,\delta,I,F)$ be an $l$-valued automaton
with $\varepsilon$-moves. Then the recognizability $rec_{\cal A}$ is
also defined as an $l$-valued unary predicate over $\sa$, and it is
given by

$rec_{\cal A}(\omega)=(\exists n\geq 0)(\exists \tau_1\in \Sigma\cup
\{\varepsilon\})\cdots (\exists
\tau_n\in\Sigma\cup\{\varepsilon\}).(\exists q_0\in Q)\cdots(\exists
q_n\in Q). (q_0\in I\wedge q_n\in F\wedge (q_0,\tau_1,q_1)\in
\delta\wedge\cdots\wedge (q_{n-1},\tau_n,q_n)\in \delta\wedge
\tau_1\cdots\tau_n=\omega)$

\noindent for all $\omega\in \sa$. The defining equation of
$rec_{\cal A}$ may be rewritten in terms of truth value as follows:

$[rec_{\cal A}
(\omega)]=\bigvee\{I(q_0)\wedge\delta(q_0,\tau_1,q_1)\wedge\cdots
\wedge\delta(q_{n-1},\tau_n,q_n)\wedge F(q_n): n\geq 0$,
$\tau_1,\cdots, \tau_n\in\Sigma\cup \{\varepsilon\}$ satisfying
$\tau_1\cdots\tau_n=\omega$, and $q_0,\cdots,q_n\in Q\}$.

We shall show that $l$-VFA and $l$-VFA$_{\varepsilon}$ are
equivalent in the sequel. First, we study a special kind of
$l$-VFA$_{\varepsilon}$ in which quantum transition is crisp, that
is, $\delta$ is a crisp subset of $Q\times
(\Sigma\cup\{\varepsilon\})\times Q$. In this case, $\delta$ can be
seen as a mapping from $Q\times (\Sigma\cup\{\varepsilon\})$ to
$2^Q$.

Let ${\cal A}=(Q,\Sigma,\delta,q_0,F)$ be an $l$-VFA$_{\varepsilon}$
with crisp quantum transition and with a unique initial state
$q_0\in Q$, the explicit expression of $rec_{\cal A}$ can be induced
as follows. First, we give the extension $\ds: 2^Q\times \sa\rw 2^Q$
using the notion of $\varepsilon$-closure. For $q\in Q$, the
$\varepsilon$-closure of $q$, denoted $EC(q)$, is defined as,

$EC(q)=\{p\in Q: $ there exists $n\geq 0$ and $q_0,\cdots,q_n$
satisfying $q_i\in\delta(q_{i-1},\varepsilon)$ for any $i=1, \cdots,
n$, in which $q_0=q$ and $q_n=p\}$.

\noindent For any subset $X$ of $Q$, the $\varepsilon$-closure of
$X$, denoted $EC(X)$, is defined as

$EC(X)=\bigcup_{q\in X}EC(q)$.

\noindent In particular, $EC(\{q\})=EC(q)$. Then $\ds$ is defined
inductively as,

$\ds(q,\varepsilon)= EC(q)$,

$\ds(q,\omega\sigma)=EC(\delta(\ds(q,\omega),\sigma))$ for any $q\in
Q$, $\omega\in \sa$ and $\sigma\in \Sigma$.

\noindent Then

$\ds(X,\omega)=\bigcup_{q\in X}\ds(q,\omega)$.

\noindent It follows that

$\ds(q,\omega\sigma)=\ds(\ds(q,\omega),\sigma)$

\noindent for any $q\in Q$, $\omega\in\sa$ and $\sigma\in \Sigma$.
By the definition of unitary predicate $rec$ over $\sa$, the truth
valued $rec_{\cal A}$ for an $l$-VFA$_{\varepsilon}$ with crisp
quantum transition is defined as follows: for any $\omega\in\sa$,

$[rec_{\cal A}](\omega)=\bigvee\{F(q):q\in \ds(q_0,\omega)\}$.

We construct an equivalent $l$-VFA ${\cal B}$ from the above ${\cal
A}$ as follows, where ${\cal B}=(Q,\Sigma,\eta,q_0,E)$. The quantum
transition $\eta$ is defined as: for any $q\in Q$ and $\sigma\in
\Sigma$,

$\eta(q,\sigma) =\ds(q,\sigma)$.

\noindent If $q\not= q_0$, then

$E(q)=F(q)$,

\noindent and

$E(q_0)=\bigvee\{F(q): q\in EC(q_0)\}$.

\noindent Note that ${\cal B}$ has no $\varepsilon$-transitions.

\begin{lemma}\label{le:e-moves1}

For any $l$-VFA$_{\varepsilon}$ with crisp quantum transition ${\cal
A}$, the $l$-VFA ${\cal B}$ constructed as above is equivalent to
${\cal A}$, i.e., $rec_{\cal A}=rec_{\cal B}$.

\end{lemma}

\noindent {\bf Proof}\ \ We wish to show by induction on $|\omega|$
that $\eta^{\ast}(q,\omega)=\ds(q,\omega)$. However, this statement
may not be true for $\omega=\varepsilon$, since
$\es(q,\varepsilon)=\{q\}$, while $\ds(q,\varepsilon)=EC(q)$. We
therefore begin our induction at 1.

Let $|\omega|=1$. Then $\omega$ is a symbol $\sigma$, and
$\eta(q,\sigma)=\ds(q,\sigma)$ by definition of $\eta$. Suppose that
the hypothesis holds for inputs of length $n$ or less. Let
$\omega=x\sigma$ be a string of length of $n+1$ with symbol $\sigma$
in $\Sigma$. Then

$\es(q,x\sigma)=\eta(\es(q,x),\sigma)$.

By the inductive hypothesis, $\es(q,x)=\ds(q,x)$. Let $\ds(q,x)=X$,
we must show that $\eta(X,\sigma)=\ds(q,x\sigma)$. But

$\eta(X,\sigma)=\bigcup_{q\in Q}\eta(q,\sigma)=\bigcup_{q\in
X}\ds(q,\sigma)$.

Then as $X=\ds(q,x)$ we have

$\bigcup_{q\in X}\ds(q,\sigma)=\ds(q,x\sigma)$.

Thus

$\es(q,x\sigma)=\ds(q,x\sigma)$.

To complete the proof we shall show that $[rec_{\cal
B}(\omega)]=\bigvee\{F(q): q\in\ds(q_0,\omega)\}$.

If $\omega=\varepsilon$, this statement is immediate from the
definition of $E$. That is, $\es(q_0,\varepsilon)=\{q_0\}$, then
$[rec_{\cal B}(\varepsilon)]=\bigvee\{E(q): q\in
\es(q_0,\varepsilon)\}=E(q_0)=\bigvee\{F(q): q\in
\ds(q_0,\varepsilon)\}$.

If $\omega\not=\varepsilon$, then $\omega=x\sigma$ for some symbol
$\sigma$. We have two cases to discuss.

Case I: $q_0\not\in \es(q_0,x\sigma)$. By the definition of $E$ and
the equality $\es(q_0,x\sigma)=\ds(q_0,x\sigma)$, it follows that

$[rec_{\cal B}(x\sigma)]=\bigvee\{E(q): q\in
\es(q_0,x\sigma)\}=\bigvee\{F(q): q\in\ds(q_0,x\sigma)\}$.

Case II: $q_0\in \es(q_0,x\sigma)$. Then $EC(q_0)\subseteq
\ds(q_0,x\sigma)=\es(q_0,x\sigma)$. Thus,

$[rec_{\cal B}(x\sigma)]=\bigvee\{E(q): q\in
\es(q_0,x\sigma)\}=\bigvee\{E(q): q\in
\ds(q_0,x\sigma)\}=\bigvee\{E(q): q\in
\ds(q_0,x\sigma)-\{q_0\}\}\vee E(q_0)=\bigvee\{F(q): q\in
\ds(q_0,x\sigma)-\{q_0\}\}\vee \bigvee\{F(q): q\in
EC(q_0)\}=\bigvee\{F(q): q\in \ds(q_0,x\sigma)\}$.

Hence, for any $\omega\in\sa$, $[rec_{\cal
B}(\omega)]=\bigvee\{F(q): q\in \ds(q_0,x\sigma)\}=[rec_{\cal
A}(\omega)]$. This shows that $rec_{\cal A}=rec_{\cal B}$, and thus
${\cal A}$ and ${\cal B}$ are equivalent. \hfill$\Box$

Let ${\cal A}=(Q,\Sigma,\delta,I,F)$ be an $l$-VFA$_{\varepsilon}$.
We construct an equivalent $l$-VFA$_{\varepsilon}$ ${\cal
B}=(P,\Sigma,\eta,S,E)$ with crisp quantum transition from ${\cal
A}$ as follows.

Let $X=Im(\delta)\cup Im(I)\cup Im(F)$, and $l_1=X_{\wedge}$. Choose
$P=2^{Q\times(l_1-\{0\})}$, and $S=\{(q,I(q)): q\in Q$ and
$I(q)\not=0\}$, then $P$ is a finite set and $S\in P$. The state
transition $\eta: P\times (\Sigma\cup\{\varepsilon\})\rw P$ is
defined by,

$\eta(\{(q,r)\},\tau)=\{(p,r\wedge\delta(q,\tau,p)): p\in Q$ and
$r\wedge\delta(q,\tau,p)\not=0\}$

\noindent for any $(q,r)\in Q\times(l_1-\{0\})$ and $\tau\in
\Sigma\cup\{\varepsilon\}$. We define

$\eta(Z,\tau)=\bigcup_{(q,r)\in Z}\eta(\{(q,r)\})$

\noindent for any $Z\in P$ and $\tau\in \Sigma\cup\{\varepsilon\}$.
Then $\eta$ is well defined as discussed in the quantum subset
construction from an $l$-VFA to an $l$-VDFA. The quantum final state
$E :P\rw l$ is defined as,

$E(Z)=\bigvee\{r\wedge F(q): (q,r)\in Z\}$.

\begin{lemma}\label{le:e-moves2}

For any $l$-VFA$_{\varepsilon}$ ${\cal A}=(Q,\Sigma,\delta,I,F)$,
the $l$-VFA$_{\varepsilon}$ with crisp quantum transition ${\cal B}$
constructed as above is equivalent to ${\cal A}$, i.e., $rec_{\cal
A}=rec_{\cal B}$.

\end{lemma}

\noindent {\bf Proof}\ \ The proof is very similar to that of
Theorem \ref{th:determinization}, we omit it here. \hfill$\Box$

Combining the above two lemmas, we can conclude the following theorem which shows the equivalence
between $l$-VFA$_{\varepsilon}$ and $l$-VFA.

\begin{theorem}\label{th:e-moves}

For any $l$-VFA$_{\varepsilon}$ ${\cal A}$, there is an
$l$-VFA ${\cal B}$ such that ${\cal A}$ and ${\cal B}$ are equivalent,
i.e., $rec_{\cal A}=rec_{\cal B}$.

\end{theorem}

Combining Theorem \ref{th:determinization} and Theorem
\ref{th:e-moves}, we can see the equivalence between
$l$-VFA$_{\varepsilon}$, $l$-VFA and $l$-VDFA.

\begin{corollary}\label{co:e-moves}

For any $l$-VFA$_{\varepsilon}$ ${\cal A}$, there is an
$l$-VDFA ${\cal B}$ such that ${\cal A}$ and ${\cal B}$ are equivalent,
i.e., $rec_{\cal A}=rec_{\cal B}$.

\end{corollary}

As an application of Theorem \ref{th:determinization}, we present
pumping lemma in the frame of quantum logic as follows.

\begin{proposition}\label{pro:pumping}{\rm (Pumping lemma in quantum
logic)} For an $l$-regular language $A:\sa\rw l$, there exists
positive integer $n$, for any input string $z\in\sa$, if $|z|\geq
n$, then there are $u,v,w\in \sa$ such that $|uv|\leq n$,
$v\not=\varepsilon$, $z=uvw$, and for any non-negative integer $l$,
the equality $A(uv^lw)= A(uvw)$ holds.

\end{proposition}

\noindent {\bf Proof}\ \ Since $A$ is $l$-regular, it is accepted by
an $l$-VFDA ${\cal A}=(Q,\Sigma,\delta,q_0,F)$ with some particular
number of states, say $n$. Consider an input of $n$ or more symbols
$z=\sigma_1\cdots\sigma_m$, $m\geq n$, and for $i=1,\cdots,m$, let
$\ds(q_0,\sigma_1\cdots\sigma_i)=q_i$. It is not possible for each
of the $n+1$ states $q_0,\cdots,q_n$ be different, since there are
only $n$ different states. Thus there are two integers $j$ and $k$,
$0\leq j<k\leq n$, such that $q_j=q_k$. Let
$u=\sigma_1\cdots\sigma_j$, $v=\sigma_{j+1}\cdots\sigma_{k}$,
$w=\sigma_{k+1}\cdots\sigma_m$, then $|uv|=k\leq n$,
$v\not=\varepsilon$ and $z=uvw$. Observing that
$\ds(q_0,\sigma_1\cdots\sigma_j\sigma_{k+1}\cdots\sigma_m)=\ds(\ds(q_0,\sigma_1
\cdots\sigma_j),\sigma_{k+1}\cdots\sigma_m)=\ds(q_j,\sigma_{k+1}\cdots\sigma_m)
=\ds(q_k,\sigma_{k+1}\cdots\sigma_m)=q_m$, and for any $l\geq 1$,
$\ds(q_0,\sigma_1\cdots\sigma_j(\sigma_{j+1}\cdots\sigma_{k})^l
\sigma_{k+1}\cdots\sigma_m)=\ds(\ds(\ds(q_0,\sigma_1\cdots\sigma_j),
(\sigma_{j+1}\cdots\sigma_{k})^l),\sigma_{k+1}\cdots\sigma_m)=\ds(\ds(q_j,
(\sigma_{j+1}\cdots\sigma_{k})^l),\sigma_{k+1}\cdots\sigma_m)
=\ds(q_k,\sigma_{k+1}\cdots\sigma_m)=q_m$. Therefore, for any $l\geq
0$, $A(uv^lw)$ = $[rec_{\cal A}(uv^lw)]$ =
$F(\ds(q_0,uv^lw))=F(q_m)=F(\ds(q_0,uvw))=[rec_{\cal
A}(uvw)]=A(uvw)$. \hfill$\Box$


\begin{remark}\label{re:quantum logic}

{\rm Lemma \ref{le:e-moves1}, Lemma \ref{le:e-moves2}, Theorem
\ref{th:e-moves}, Corollary \ref{co:e-moves} and Proposition
\ref{pro:pumping} (and all propositions in Section 3) can be
restated in the language of quantum logic, as done in Theorem
\ref{th:determinization}, we left them to the readers which are
interested in stating the related propositions in logic language.}

\end{remark}

\section{Kleene Theorem for $l$-valued finite automata}

We use $lR(\Sigma)$ to denote the set of $l$-regular languages over
$\Sigma$. Up to now, we still do not know whether $lR(\Sigma)$ is
closed under the operations of meet, complement and Kleene closure
of $l$-valued regular languages. Indeed, in \cite{ying05}, Ying gave
some conditions using the notion of commutators to guarantee
$lR(\Sigma)$ being closed under the above mentioned operations.
Since the above mentioned restrictions, Kleene theorem for $l$-VFA
depends on the notion of commutators. We shall show that all these
restrictions are not necessary in this section. In fact, we shall
show that $lR(\Sigma)$ is closed under the operations of meet,
complement and Kleene closure of $l$-valued regular languages.
Furthermore, Kleene theorem holds in the frame of quantum logic.

Let us recall the operations of $l$-valued languages
(\cite{ying05}): for $A,B\in l(\sa)$ and $r\in l$, the union $A\vee
B$, the intersection $A\wedge B$, the complement $A^{\bot}$, the
scalar product $rA$, the concatenation $AB$, the Kleene closure
$A^{\ast}$ are defined as follows: for any $\omega\in\sa$, $A\vee
B(\omega)=A(\omega)\vee B(\omega)$, $A\wedge B(\omega)=
A(\omega)\wedge B(\omega)$, $A^{\bot}(\omega)=A(\omega)^{\bot}$,
$rA(\omega)=r\wedge A(\omega)$,
$AB(\omega)=\bigvee\{A(\omega_1)\wedge B(\omega_2):
\omega_1\omega_2=\omega\}$,
$A^{\ast}(\omega)=\bigvee\{A(\omega_1)\wedge\cdots\wedge
A(\omega_n): n\geq 0, \omega_1\cdots\omega_n=\omega\}$.

We first give a structure characterization of $l$-valued regular
languages.

\begin{theorem}\label{thm:regular}

Let $A: \sa\rw l$ be an $l$-valued language over $\Sigma$. Then the
following statements are equivalent.

(1) $A$ is an $l$-regular language.

(2) There exist $k_1,\cdots,k_m\in l-\{0\}$, and regular languages
$L_1, \cdots, L_m$ such that $A=\bigvee^n_{i=1}k_i 1_{L_i}$, where
$1_{L_i}$ denotes the characteristic function of $L_i$.

(3) There exist $k_1,\cdots,k_m\in l-\{0\}$, and pairwise disjoint
regular languages $L_1, \cdots, L_m$ satisfying the equality
$A=\bigvee^n_{i=1}k_i 1_{L_i}$.

\end{theorem}

\noindent {\bf Proof}\ \ (1)$\Longrightarrow$(3)\ \ Since $A$ is an
$l$-valued regular language, there is an $l$-VDFA ${\cal
A}=(Q,\Sigma,\delta,q_0,F)$ recognized $A$. That is, for all
$\omega\in\sa$, $A(\omega)=[rec_{\cal
A}(\omega)]=F(\ds(q_0,\omega))$. Write
$Im(F)-\{0\}=\{k_1,\cdots,k_m\}$, and let $F_i=\{q\in Q:
F(q)=k_i\}$, For this $F_i$, we construct a DFA, ${\cal
A}_i=(Q,\Sigma,\delta,q_0, F_i)$. Let the language recognized by
${\cal A}_i$ be $L_i$, then $L_i$ is a regular language, and
evidently, the family $\{L_1,\cdots,L_m\}$ is pairwise disjoint.
Moreover, $A(\omega)=r$ iff $F(\ds(q_0,\omega))=r$, iff there is $i$
such that $r=k_i$ and $\omega\in L_i$, which shows that
$A=\bigvee^m_{i=1}k_i 1_{L_i}$.

(3)$\Longrightarrow$(2) is obvious.

(2)$\Longrightarrow$(1)\ \ Since each $L_i$ is regular, there is a
DFA ${\cal A}_i=(Q_i,\Sigma,\delta,q_{0i},F_i)$ recognized $L_i$. We
can assume that $Q_i\cap Q_j=\emptyset$ whenever $i\not=j$. Define
an $l$-VFA, ${\cal A}=(Q, \Sigma, \delta, q_0, F)$ as follows,
$Q=\bigcup_{i=1}^mQ_i\cup\{q_0\}$, where $q_0\not\in \bigcup_{i=1}^m
Q_i$, and $\delta: Q\times \Sigma\rw 2^Q$ is,
$\delta(q_0,\sigma)=\{\delta_1(q_{01},\sigma),\cdots,\delta_m(q_{0m},\sigma)\}$,
for $q\in Q_i$, $\delta(q,\sigma)=\delta_i(q,\sigma)$;
$F(q_0)=\bigvee\{k_i: q_{0i}\in F_i\}$, and when $q\not=q_0$,
$$F(q)=\left\{
\begin{array}{lll}
k_i, & $ if $ q\in F_i\\
0, & $ otherwise$.
\end{array}
\right.$$ \noindent Then it can be easily verified that $A=rec_{\cal
A}=\bigvee^m_{i=1}k_i 1_{L_i}$. Hence $A$ is an $l$-valued regular
language. \hfill$\Box$

We call the $l$-valued language satisfying the condition (2) or (3)
in the above theorem the $l$-valued recognizable step language, and
write the set of all $l$-valued recognizable languages on $\Sigma$
as $step(\Sigma)$, which is equal to $lR(\Sigma)$.

The following proposition gives the level characterization of
$l$-valued recognizable step languages.

\begin{corollary}\label{co:regular}

Let $A: \sa\rw l$ be an $l$-valued language over $\Sigma$. Then the
following statements are equivalent.

(1) $A$ is an $l$-regular language.

(2) The image set $Im(A)$ is finite, and for any $r\in Im(A)-\{0\}$,
the $r$-cut of $A$, $A_r=\{\omega\in \sa: A(\omega)\geq r\}$ is a
regular language on $\Sigma$ and $A=\bigvee_{r\in
Im-\{0\}}r1_{A_r}$.

(3) The image set $Im(A)$ is finite, and for any $r\in Im(A)-\{0\}$,
the $r$-level of $A$, $A_{[r]}=\{\omega\in \sa: A(\omega)= r\}$ is a
regular language on $\Sigma$ and $A=\bigvee_{r\in
Im-\{0\}}r1_{A_{[r]}}$.

\end{corollary}

\begin{theorem}\label{thm:operation}

The family $step(\Sigma)$ or $lR(\Sigma)$ is closed under the
operations of union, intersection, scalar product, complement,
concatenation and Kleene closure.

\end{theorem}

\noindent {\bf Proof}\ \ Let $A,B\in step(\Sigma)$. By Theorem
\ref{thm:regular}, we can assume $A=\bigvee_{i=1}^m k_i 1_{L_i}$,
$B=\bigvee_{j=1}^n d_j 1_{M_j}$, where, all $L_i$ and $M_j$ are
regular languages and $\{L_i\}_{i=1}^m$ are pairwise disjoint,
$\{M_j\}_{i=1}^m$ are also pairwise disjoint.

With respect to the union,  we have $A\vee B=\bigvee_{i=1}^m k_i
1_{L_i}\vee \bigvee_{j=1}^n d_j 1_{M_j}$. By Theorem
\ref{thm:regular}, it follows that $A\vee B \in step(\Sigma)$.

With respect to the intersection, we have $A\wedge
B=\bigvee_{i=1}^m\bigvee_{j=1}^n (k_i\wedge d_j)1_{L_i\cap M_j}$. By
Theorem \ref{thm:regular}, it follows that $A\wedge B \in
step(\Sigma)$.

With respect to the scalar product, for each $r\in l$, we have $rA(\omega)=r\wedge A(\omega)$,
then $rA=\bigvee_{i=1}^m (r\wedge k_i) 1_{L_i}$. Therefore, $rA\in step(\Sigma)$.

For the complement operation, since
$A^{\bot}(\omega)=A(\omega)^{\bot}$, it follows that
$A^{\bot}=\bigvee_{i=1}^m k_i^{\bot} 1_{L_i}\vee
1_{\sa-(L_1\cup\cdots\cup L_m)}$. By Theorem \ref{thm:regular}, it
follows that $A^{\bot}\in step(\Sigma)$.

For the operation of concatenation, since
$AB(\omega)=\bigvee\{A(\omega_1)\wedge B(\omega_2):
\omega=\omega_1\omega_2\}$, it follows that
$AB=\bigvee_{i=1}^m\bigvee_{j=1}^n (k_i\wedge d_j)1_{L_iM_j}$. This
shows that $AB\in step(\Sigma)$.

For the Kleene closure, $A^{\ast}$ is defined by, $A^{\ast}(\omega)=
\bigvee\{A(\omega_1)\wedge \cdots A(\omega_k): k\geq 0, \omega=
\omega_1\cdots\omega_k\}$ for any $\omega\in \sa$. Since
$A=\bigvee_{i=1}^m k_i 1_{L_i}$, and $L_1,\cdots, L_m$ are pairwise
disjoint regular languages and $k_i\not=0$ for each $i$, it follows
that $Im(A)-\{0\}=\{k_1,\cdots,k_m\}$, and $L_i=\{\omega\in\sa:
A(\omega)=k_i\}$ ($i=1,\cdots, m$). For any nonempty subset $K$ of
the set $\{1,2,\cdots,m\}$, we can assume that $K=
\{i_1,\cdots,i_s\}$. Let $r_K=r_{i_1}\wedge\cdots \wedge r_{i_s}$,
$L(K)= \bigcup_{p_1\cdots p_s}L_{p_1}^+ L_{p_2}^+ L_{p_1}^{\ast}
L_{p_3}^+ (L_{p_1}\cup L_{p_2})^{\ast} \cdots L_{p_{s-1}}^+
(L_{p_1}\cup\cdots \cup L_{p_{s-2}})^{\ast} L_{p_s}^+
(L_{p_1}\cup\cdots\cup L_{p_s})^{\ast}$, where $p_1\cdots p_s$ is a
permutation of $\{i_1,\cdots,i_s\}$, and $L(K)$ is taken unions
under all permutations of $\{i_1,\cdots,i_s\}$. Hence $L(K)$ is a
regular language. It is easily verified that
$A^{\ast}=\bigvee_{\emptyset\not=K\subseteq \{1,2,\cdots,m\}} r_K
1_{L(K)}\vee 1_{\{\varepsilon\}}$. By Theorem \ref{thm:regular}, it
follows that $A^{\ast}\in step(\Sigma)$. \hfill$\Box$

\begin{definition}\label{def:expression}{\rm \cite{ying05}}
{\rm The language of {\sl $l$-valued regular expressions over
alphabet $\Sigma$} has the alphabet
$(\Sigma\cup\{\varepsilon,\emptyset\})\cup (l\cup
\{+,\cdot,\ast\})$. The symbols in
$\Sigma\cup\{\varepsilon,\emptyset\}$ will be used to denote atomic
expressions, and the symbols in $l\cup \{+,\cdot,\ast\}$ will be
used to stand for operators for building up compound expressions:
$\ast$ and all $r\in l$ are the unary operators, and $+,\cdot$ are
binary ones. We use $\alpha,\beta$ to act as meta-symbols for
regular expressions and $L(\alpha)$ for the language denoted by
expression $\alpha$. More explicitly, $L(\alpha)$ will be used to
denote an $l$-valued subset of $\sa$; that is, $L(\alpha)\in
l^{\sa}$. The $l$-valued regular expressions and the $l$-valued
languages denoted by them are formally defined as follows:

(1) For each $\sigma\in \Sigma$, $\sigma$ is a regular expression, and
$L(\sigma)=\{\sigma\}$; $\varepsilon$ and $\emptyset$
are regular expressions, and $L(\varepsilon)=\{\varepsilon\}$, $L(\emptyset)=\emptyset$.

(2) If both $\alpha$ and $\beta$ are regular expressions, then for
each $r\in l$, $r\alpha$, $\alpha+\beta$, $\alpha\cdot \beta$,
$\alpha^{\ast}$ are all regular expressions, and
$L(r\alpha)=rL(\alpha)$, $L(\alpha+\beta)=L(\alpha)\vee L(\beta)$,
$L(\alpha\cdot \beta)=L(\alpha)L(\beta)$,
$L(\alpha^{\ast})=L(\alpha)^{\ast}$.}

\end{definition}

\begin{theorem}\label{thm:kleene}{\rm (Kleene Theorem in quantum
logic)} For an $l$-valued language $A\in l(\sa)$, $A$ can be
recognized by an $l$-VFA iff there exists an $l$-valued regular
expression $\alpha$ over $\Sigma$ such that $A=L(\alpha)$.

\end{theorem}

\noindent {\bf Proof}\ \ If $A$ can be recognized by an $l$-VFA,
then by Theorem \ref{thm:regular}, there exist $k_1,\cdots,k_n\in
l-\{0\}$, and regular languages $L_1, \cdots, L_n$ such that
$A=\bigvee^n_{i=1}k_i 1_{L_i}$. Since each $L_i$ is a regular
language, by classical Kleene Theorem, there exists a regular
expression $\alpha_i$ over $\Sigma$ such that $L(\alpha_i)=L_i$. Let
$\alpha=k_1\alpha_1+\cdots+k_n \alpha_n$, then $\alpha$ is an
$l$-valued regular expression, and $L(\alpha)=\bigvee_{i=1}^n
k_iL(\alpha_i) =\bigvee^n_{i=1}k_i 1_{L_i}=A$.

Conversely, assume that there exists an $l$-valued regular
expression $\alpha$ such that $A=L(\alpha)$. We show that $A$ can be
recognized by an $l$-VFA inductively on the number of operation
symbols occurring in $\alpha$. If there is no operation symbol in
$\alpha$, then $\alpha=\sigma\in \Sigma, \varepsilon$ or
$\emptyset$. In this case, $L(\alpha)=\{\sigma\}, \{\varepsilon\}$
or $\emptyset$, and $L(\alpha)$ can be recognized by a classical
DFA. The classical DFA is evidently an $l$-VDFA, so $L(\alpha)$ can
be recognized by an $l$-VDFA in this case. Inductively, since the
family of recognizable languages by $l$-VDFA is closed under union,
intersection, scalar product, concatenation and Kleene closure (by
Theorem \ref{thm:operation}), it follows that $L(\alpha)$ can be
recognized by an $l$-VDFA for any $l$-valued regular expression
$\alpha$. \hfill$\Box$

\section  {Conclusion}

In this paper, we introduced the quantum subset construction of
orthomodular lattice-valued finite automata, then we proved the
equivalence between orthomodular lattice-valued finite automata,
orthomodular lattice-valued deterministic finite automata and
orthomodular lattice-valued finite automata with
$\varepsilon$-moves. We give a simple characterization of
orthomodular lattice-valued languages recognized by orthomodular
lattice-valued finite automata, then we proved that the Kleene
theorem holds in the frame of quantum logic, many results in
\cite{ying05} can be strengthen such as the pumping lemma in the
frame of quantum logic using the results of this paper.


\begin{thebibliography}{99}
\baselineskip 15pt

\bibitem{ambainis02} A.Ambainis, J.Watrous, Two-way finite automata
with quantum and classical states, Theoretical Computer Science,
287(2002), 299-311.

\bibitem{eilenberg74} S.Eilenberg,  Automata, Languages and Machines, vol. A, vol B,
Academic Press, New Yok, 1974.

\bibitem{esik07} Z. \'{E}sik, W. Kuich, Modern Automata Theory, 2007,
see http://dmg.tuwien.ac.at/kuich/.

\bibitem{gruska99}J. Gruska, Quantum Computing, McGraw-Hill, London, 1999.

\bibitem{hopcroft79} J.E. Hopcroft ,  J. D. Ullman,
Introduction to Automata Theory,
 Languages and Computation, Addison-Wesley, New York, 1979

\bibitem{kalmbach83} G. Kalmbach, Orthomodular Lattices,
Academic Press, London, 1983.

\bibitem{khoussainov01} B. Khoussainov, A. Nerode,  Automata Theory and its
Applications, Birk\"{a}user, Boston, 2001.

\bibitem{kleene56}S.C. Kleene, Representation of events in nerve nets and finite automata,
in: Automata Studies, ed. by C.E. Shannon and J. McCarthy, Princeton University Press,
Princeton, NJ, 1956, 3-42.

\bibitem{li93} Y.M. Li, Z.H. Li, Free semilattices and strongly free semilattices
generated by partially ordered sets, Northeastern Mathematical
Journal, 9(3)(1993), 359-366.

\bibitem{li05}Y.M. Li, W.Pedrycz, Fuzzy finite automata and
 fuzzy regular expressions with membership values in
 lattice-ordered monoids, Fuzzy Sets and Systems, 156(2005), 68-92.

\bibitem{moore00} C.Moore, J.P. Crutchfield, Quantum automata and
quantum grammars, Theoretical Computer Science, 237(2000), 275-306.

\bibitem{nc00} M.A.Nielsen,  I.L. Chuang, Quantum Computation and Quantum
Information, Cambridge University, Cambridge, 2000.

\bibitem{qiu04} D.W. Qiu, Automata theory based on qunatum logic:
some characterizations, Information and Computation, 190(2004),
179-195.

\bibitem{qiu071} D.W. Qiu, Automata theory based on qunatum logic: reversibilities
and pushdown automata, Theoretical Computer Science, 386(2007),
38-56.

\bibitem{qiu072} D.W. Qiu, Notes on automata theory based on quantum logic,
Science in China Series F: Information Sciences, 50(2)(2007),
154-169.

\bibitem{rabin59} M.O. Rabin, D. Scott, Finite automata and their decision problems,
IBM J. Research and Development, 3(1959), 114-125.

\bibitem{ying001} M.S. Ying, Automata theory based on quantum logic (I), International Journal of
Theoretical Physics, 39(2000), 981-991.

\bibitem{ying002} M.S. Ying,  Automata theory based on quantum logic (II), International Journal of
Theoretical Physics, 39(2000), 2545-2557.

\bibitem{ying05} M.S. Ying,  A theory of computation based on quantum logic (I),
Theoretical Computer Science, 344(2005), 134-207.



\end{thebibliography}
\end{document}